%% file: Kepler91b_RVconfirmation.tex
\documentclass[letter]{aa} % for the letters 
\usepackage{graphicx}
\usepackage{graphicx}
\usepackage{lscape}
\usepackage{longtable}
\usepackage{natbib}
\usepackage{color}
\usepackage{array} 
\bibpunct{(}{)}{;}{a}{}{,} % to follow the A&A style
\usepackage{txfonts}
\begin{document}

   \title{Radial velocity confirmation of Kepler-91 b}

   \subtitle{Additional evidence of its planetary nature using the Calar Alto/CAFE instrument}

   \author{J. Lillo-Box\inst{1}, D. Barrado\inst{1}, Th. Henning\inst{2}, L. Mancini\inst{2}, S. Ciceri\inst{2}, P.~Figueira\inst{3}, \\N.C.~Santos\inst{3,4}, J.~Aceituno\inst{5}, S.~S\'anchez\inst{6}
          }

%1
  \institute{Depto. de Astrof\'isica, Centro de Astrobiolog\'ia (CSIC-INTA), ESAC campus 28691 Villanueva de la Ca\~nada (Madrid), Spain\\
              \email{Jorge.Lillo@cab.inta-csic.es}\and
%2
Max Planck Institute for Astronomy, K\"onigstuhl 17, 69117 Heidelberg, Germany  \and %\and
%3
Centro de Astrof\'{i}sica, Universidade do Porto, Rua das Estrelas, 4150-762 Porto, Portugal\and
%4
Departamento de F\'{i}sica e Astronomia, Faculdade de Ci\^{e}ncias, Universidade do Porto, Portugal \and
%5
Centro Astron\'omico Hispano-Alem\'an (CAHA). Calar Alto Observatory, c/ Jes\'us Durb\'an Rem\'on 2-2, 04004, Almer\'ia, Spain. \and
%6
Instituto de Astronom\'ia,Universidad Nacional Auton\'oma de M\'exico, A.P. 70-264, 04510, M\'exico,D.F.               
            }

   \date{Accepted on July 31st, 2014}

\titlerunning{Radial velocity confirmation of Kepler-91 b}
\authorrunning{Lillo-Box et al.}

% \abstract{}{}{}{}{} 
% 5 {} token are mandatory
 
  \abstract
   {The object transiting the star \object{Kepler-91} was recently assessed as being of planetary nature. The confirmation was achieved by analysing the light-curve modulations observed in the \textit{Kepler} data. However, quasi-simultaneous studies claimed a self-luminous nature for this object, thus rejecting it as a planet. In this work, we apply an { independent} approach to confirm the planetary mass of  \object{Kepler-91b} by using multi-epoch high-resolution spectroscopy obtained with the Calar Alto Fiber-fed Echelle spectrograph (CAFE). We obtain the physical and orbital parameters with the radial velocity technique. In particular, we derive a value of $1.09 \pm 0.20\,M_{\mathrm{Jup}}$ for the mass of Kepler-91b, in excellent agreement with our previous estimate that was based on the orbital brightness modulation.}


   \maketitle
%
%________________________________________________________________

\section{Introduction}

Several techniques, some previously applied to binary systems, have been recently improved to obtain the mass of planetary-size objects. They include radial velocity (RV) measurements \citep{mayor95,santerne12}, accurate astrometry \citep{muterspaugh10,lazorenko11}, microlensing \citep{mao91,gould92}, transit-timing variations \cite[e.g.][]{holman10}, or light-curve modulations \citep[e.g.][]{borucki09}. Because of its easy applicability to a wide range of masses, the RV method has been the most popular in the past two decades. Currently, thanks to the unprecedented  accurate photometry obtained by the {\it Kepler} telescope, several studies have confirmed the light-curve modulations produced by extrasolar planets \citep[e.g.][]{borucki09, shporer11, quintana13}. The modelling of these modulations can provide the mass of the bound object, which makes it an alternative method with to confirm its planetary nature. 

\cite{lillo-box14} used this photometric technique (together with a careful characterisation of the host star) to establish the planetary nature of the object transiting the star Kepler-91 (KOI-2133 or KIC 8219268). It was detected by the {\it Kepler} mission \citep{borucki11} and announced as a planet candidate in the second release of the mission on February 2012 \citep{batalha13}. The exhaustive analysis of the light-curve modulations, transit dim, and asteroseismic signals performed by \cite{lillo-box14} led us to classify this planet as an inflated hot Jupiter ($M_p = 0.88^{+0.17}_{-0.33}$ and $R_p=1.384^{+0.011}_{-0.054}$) that orbits very close ($r=2.32^{+0.07}_{-0.22} R_{\star}$ at periastron passage) to a K3~III giant star ($M_{\star}=1.31\pm0.10\ M_{\odot}$, $R_{\star}=6.30\pm0.16\ R_{\odot}$). 

However, two studies \citep{esteves13,sliski14} claimed a possible non-planetary nature for this system. Both papers suggested that it might instead be a self-luminous (and thus star-like) object. Based on transit-fitting analysis and a claimed detection of the secondary eclipse, \cite{esteves13} concluded that the object emits light so they rejected it as a planet. In contrast, \cite{angerhausen14} provided a high-level analysis of the entire light-curve (encompassing all available {\it Kepler} observations). They detected several dimmings in the phase-folded light-curve, achieved similar conclusions as \cite{lillo-box14}, and suggested that the detected dimmings are poorly explained with a secondary eclipse alone. \cite{sliski14} estimated (also with transit-fitting) the mean stellar density of the star and compared it with the asteroseismic determination. The significant difference found between the two values was attributed to the non-planetary nature of the transiting object. 

Therefore, the problem is still open. In this paper, we provide additional and independent confirmation of the planetary nature of Kepler-91b.

%__________________________________________________________________
\section{Observations and data analysis \label{sec:observations}}

\subsection{Observations and basic reduction}

High-resolution spectra were collected during the {\it Kepler} observing window of May-July 2012 with the Calar Alto Fiber-fed Echelle spectrograph \citep[CAFE, ][]{aceituno13}, mounted on the 2.2m telescope in Calar Alto Observatory (Almer\'ia, Spain). The instrument provides an average spectral resolution of around $\lambda / \Delta \lambda = 63000$ across the whole spectral range between 4000~\AA\ and 9500~\AA, allowing nominal radial velocity accuracies at the level of few tens of meters per second for FGK stars.

The exposure time for Kepler-91, a $m_{\rm Kep}=12.5$ mag giant star (spectral type K3 III), was set between 1800 and 2700\,s, depending on the weather conditions (seeing, atmospheric transparency, etc.). We typically obtained from two to three spectra per night for this object. The signal-to-noise ratio (hereafter S/N) was calculated as the inverse of the root mean square of the spectra over a continuum region without lines { centred at 6500~\AA\ (where the efficiency peak of CAFE is located).}

In total, 40 spectra were acquired in 20 nights, with a median S/N of 11. Continuum and bias images were obtained for reduction purposes and thorium-argon (ThAr) arcs were acquired after each science spectrum to calibrate the wavelength accurately. All images were processed with the pipeline provided by the observatory \citep[see][for further details on the reduction process]{aceituno13}. It uses a specific ThAr line list of several hundred features to achieve wavelength calibration with a precision at the 1 m/s level. Every spectrum was wavelength-calibrated with the immediate arc obtained just after the science observation at the same telescope position. The data presented here were taken during the earlier periods of operations of the instrument, when the thermal and vibrational control systems were still not fully operational. 

   \begin{figure}[hbft]
     \centering 
   \includegraphics[width=0.5\textwidth]{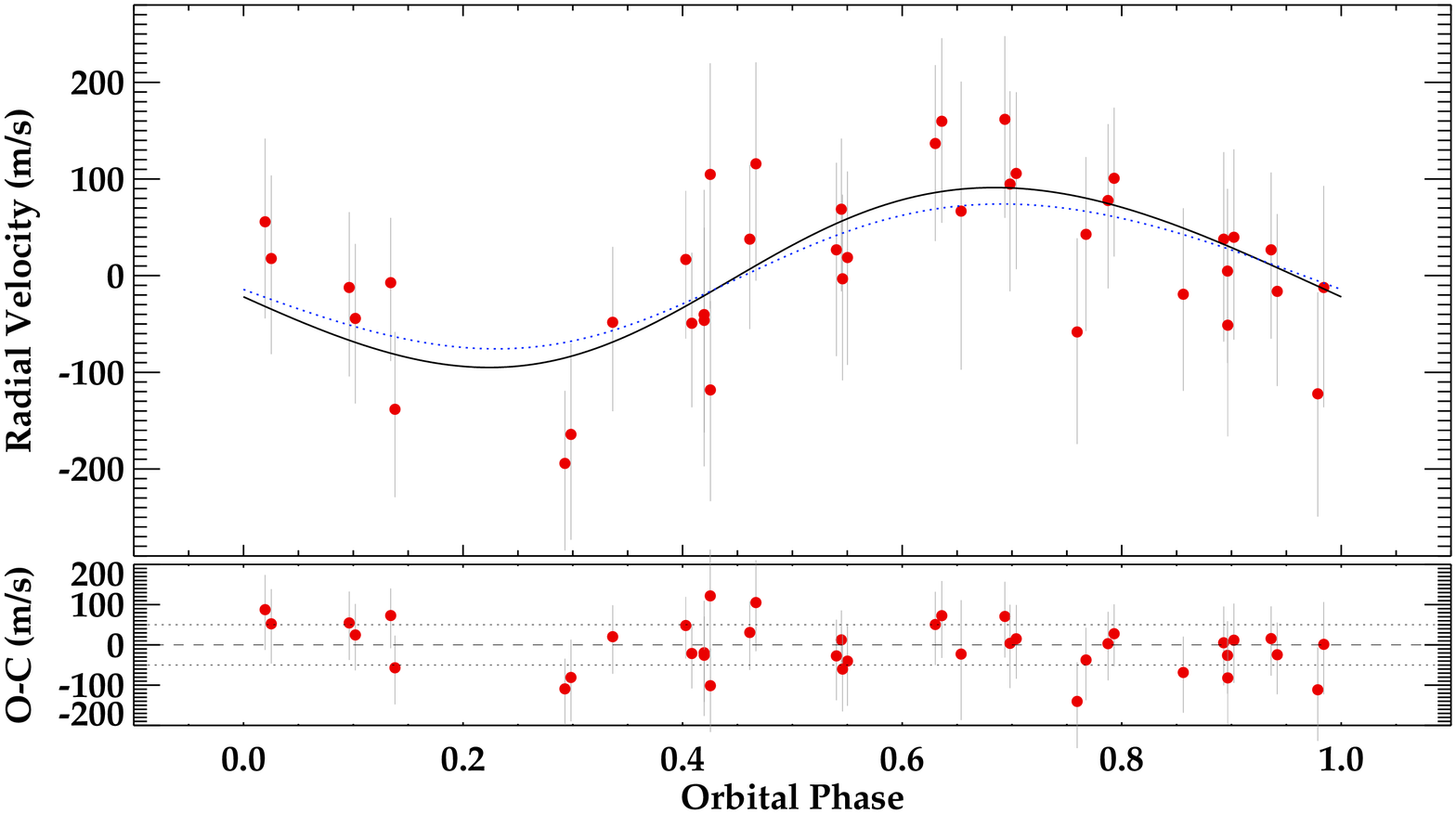}
   \caption{Radial velocity data (red circles). The solid black line shows the fit to the acquired radial velocity data by assuming the period obtained by the {\it Kepler} team \citep{batalha13} and the small eccentricity derived in \cite{lillo-box14} using the light-curve modulations (REB). The dotted line represents the independent curve obtained by using the parameters extracted from \cite{lillo-box14}, using the REB modulations.}
         \label{fig:rvcurve}
   \end{figure}

\subsection{Extracting the radial velocity with {\it GAbox}}

We applied an active cross-correlation method to extract the radial velocity information of the reduced spectra. \cite{zucker03},  it is showed that maximum-likelihood parameter determination is equivalent to cross-correlation (see Sect.~2 in that paper). The method assumes that the observed spectrum, $f(n)$, can be modelled by a template, $g(n)$, scaled by a constant ($a_0$), shifted by a determined number of bins ($s_0$) with the addition of random white Gaussian noise with a specific standard deviation ($\sigma_0$),  $f(n) = a_0 g(n-s_0)+ \mathcal{N}(0,\sigma_0^2)$. Thus, the natural logarithm of the likelihood function becomes
\begin{equation}
\label{eq:like}
\log{L} = -N\log{\sigma_0} - \frac{1}{2\sigma_0^2} \sum_n { [f(n)-a_0g(n-s_0)]^2  } + C, 
\end{equation}

\noindent where $N$ is the total number of bins (pixels) of the spectrum and $C$ is a constant independent of the parameters. The set of parameters $[\hat{s}, \hat{a}, \hat{\sigma}]$ maximising the function provides the best fit of the modified template to the observations. In particular, $\hat{s}$ can be identified with the radial velocity of the star.  This method assumes zero mean for both the template and the observed spectrum, therefore {  we subtracted their corresponding means.}

We used a modified version of our genetic algorithm that we presented in \cite{lillo-box14} (\textit{GAbox}), to find the set of parameters maximising the likelihood in Eq.~\ref{eq:like}. { We used a synthetic spectrum from \cite{coelho05} of the same spectral type as Kepler-91 as a template.} This improved algorithm searches for the best-fit set of parameters with no need of exploring the whole parameter space \citep[see general details of the method in][]{lillo-box14}. In this case, the routine first uses large step sizes of 5 km/s to search for the rough region in which the maximum is located. When this region is found, the step sizes start to decrease gradually (until the 1 m/s level) each time the algorithm finds a maximum in the likelihood function. After the final maximum has been found, the process is repeated G times, with G being the number of super-generations provided by the user (typically $G>20$). The final radial velocity of a specific order ($\hat{s}$) is obtained as the median of the G convergence values. The upper and lower confidence levels of each $G_i$ value correspond to the 3$\sigma$ statistical uncertainty. The final uncertainty of the calculated RV of the order ($\sigma_{\hat{s}}$) is computed by bootstrapping all G values with their corresponding uncertainties. 

We determined one shift per order (84 in total for CAFE) and combined them to obtain the RV of the star. The final value and its uncertainty were thus computed by the median of all orders and bootstrapping the individual results to obtain 3$\sigma$ errors.

\input{Table_Observations}

%__________________________________________________________________

\section{Results \label{sec:results}}

Table~\ref{tab:observations} summarises the observing characteristics (Julian date, exposure time, S/N, and phase) as well as the RV values for each epoch obtained as explained in Sect.~\S~\ref{sec:observations}. In Fig.~\ref{fig:rvcurve}, we show the phase-folded RV data.

\begin{figure}[hbft]
     \centering 
   \includegraphics[width=0.45\textwidth]{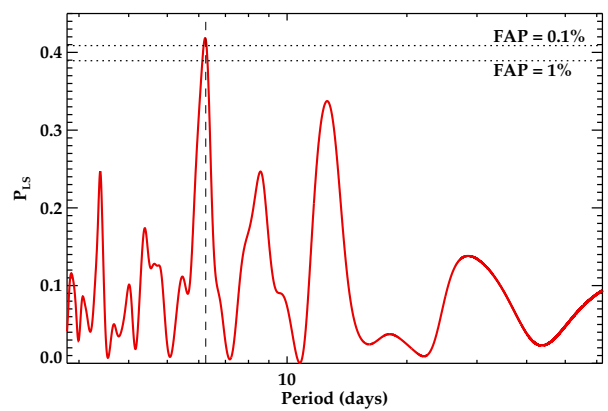}
   \caption{Lomb-Scargle periodogram of the radial velocity data obtained with CAFE. The dotted lines show the false-alarm probability levels of $FAP=0.1$\% and $FAP=1$\%. The vertical dashed line shows the period derived by transit detection. The detected peak at $6.23\pm0.03$ days in this RV periodogram has an FAP$=0.09$\%.}
         \label{fig:periodogram}
   \end{figure}

Prior to fitting a Keplerian orbit to the RV data, we performed a Lomb-Scargle periodogram (Fig.~\ref{fig:periodogram}) to check whether we detect the planetary signal at the expected period \citep[$T=6.246580 \pm 0.000082$ days as determined by][]{batalha13}. We restricted the period search to a specific range. The longest period explored was set to the longest time span between our observations (i.e. $T_{max} = t_{max}-t_{min} = 62$ days, where $t_{max}$ and $t_{min}$ are the earliest and latest Julian dates in our observations). Since the observations are unevenly separated, the shortest period searched was set to the median of the inverse time interval between data points, as was proposed by \cite{debosscher07} and \cite{ivezic13}, $T_{min}=\overline{\Delta t}=2.8$ days\footnote{\cite{eyer99} claimed that for most practical cases, lower periodicities (higher frequencies) can be detected even for strongly (but randomly) under-sampled observations.}.  The significant peak in the power spectrum (with a false-alarm probability\footnote{ Calculated by using the {\it astroML} python module \citep{vanderplas12} and its bootstrapping package.} of $FAP=0.09$\%, over the 0.1\% level) coincides with the expected period of the planet. This provides clear confirmation for the detection of a periodic signal. Consequently, we can affirm that we are detected the RV signal of  Kepler-91b.

We then used the RVLIN software\footnote{http://exoplanets.org/code/} \citep{wright09} and its additional package BOOTTRAN for parameter uncertainties estimation with bootstrapping \cite[described in][]{wang12} to fit our RV data to a Keplerian orbital solution. Since we have extensive observations of its transit, the period of the transiting object can be far more accurately determined by the transit analysis. Thus, we decided to fix the period to that provided by \cite{batalha13}. Because of the relatively large uncertainties in the RV and incomplete coverage of the RV curve, we also decided to fix the eccentricity of the orbit to the slightly non-circular  value  determined by \cite{lillo-box14}, $e=0.066^{+0.013}_{-0.017}$. The free parameters for this fitting were the semi-amplitude of the RV variations ($K$), the systemic velocity of the system ($V_{sys}$), and the orbital argument of the periastron ($\omega$).

We used the asteroseismic determination of the stellar mass of the host star by \cite{lillo-box14}, $M_{\star}=1.31\pm0.1~M_{\odot}$, to obtain an accurate value of the minimum mass of the transiting object (i.e $M_p\sin{i}$). Moreover, we  know that the orbit of this planet is highly inclined with respect to our line of sight. The inclination was also provided by \cite{lillo-box14} ($i=68.5^{+1.0}_{-2.0}$ degrees) and is supported by previous light-curve analysis such as \cite{tenenbaum12}, who derived  $i = 71.4 \pm 2.5$ degrees, in good agreement with our value. Thus, we can directly determine the absolute mass of the orbiting object. 

 The results of the RVLIN fitting process are shown in Table~\ref{tab:rvresults}, the fitted model is plotted in Fig.~\ref{fig:rvcurve}. We investigated the significance of that fit against a constant model (which would imply that we are just detecting noise). We infer a Bayesian information criterion (BIC\footnote{BIC$=\chi^2+k\log{N}$, where $\chi^2=\sum \frac{(RV_{model}-RV_{obs})^2}{\sigma_{RV}^2}$, $k$ is the number of free parameters, and $N$ is the number of data points.}) for the constant model and for the RV model of BIC$_{const.}=30.5$ and BIC$_{RV}=27.8$\footnote{ Assuming a circular orbit, we obtain BIC$_{RV(e=0)}=28.3$.}. This implies a $\Delta$BIC$=2.7$, which provides positive (although not strong) evidence for the RV model (positive detection) against the constant model (negative detection). Alternatively, we obtain a value of 12.7 for an F-test with weighted residuals\footnote{$F=\frac{\chi^2_1-\chi^2_2}{p_2-p_1} / \frac{\chi^2_2}{N-p_2}$, where $p_1$ and $p_2$ are the free parameters of both models (so that $p_2>p_1$), and $N$  the number of data points.}. This value is higher than the corresponding value of the F-distribution for a 99\% confidence level, $F_{0.01}(p_2-p_1,N-p_2) = F_{0.01}(3,36) = 4.38$. Thus, we can confirm that the detected RV variability is significant at 99\% confidence level with respect to pure noise.

  The radial velocity data confirm the Jupiter-like mass ($M_p\sin{i}=1.01\pm 0.18~M_{Jup}$) of the object orbiting Kepler-91. Using our previous value for the inclination, the absolute mass becomes $M_p = 1.09\pm 0.20~M_{Jup}$. This result agrees well, within the uncertainties, with the derived mass in the confirmation paper of  Kepler-91b \citep{lillo-box14}, $M_p = 0.88^{+0.17}_{-0.33}~M_{Jup}$. Moreover, the semi-major axis $a_{RV} = 0.0726\pm 0.0019$ AU also agrees extremely well with the value determined by the light-curve modulations in \cite{lillo-box14} of $a_{REB}=0.072^{+0.002}_{-0.007}$ AU. The corresponding radial velocity model using the photometrically derived parameters obtained by \cite{lillo-box14} is also plotted in Fig.~\ref{fig:rvcurve} for comparison purposes. 

\input{Table_RVresults}

%______________________________________________________________

\section{Discussion and conclusions \label{sec:conclusions}}

The fit to the RV data provides orbital parameters that agree excellently well with those obtained by the photometric analysis provided in the confirmation paper\footnote{ The agreement between the ellipsoidal modulation mass and the RV mass has already been demonstrated in systems such TrES-2 or HAT-P-7 \citep[see][]{faigler14}.}. Thus, we present independent support of the planetary-mass of the object transiting Kepler-91. Moreover, the transit and asteroseismic analysis yield a radius of the transiting object of $R_p = 1.384^{+0.011}_{-0.054}~R_{Jup}$, providing a mean density of $\rho_p =0.33\pm0.08~\rho_{Jup}$. 

These results disagree with \cite{esteves13} and \cite{sliski14}.  The former was already discussed in \cite{lillo-box14} (Sect.~5.2), where several arguments were given against the proposed self-luminous scenario for Kepler-91b. In brief, \cite{esteves13} derived a mass for the transiting object of  $5.92^{+0.68}_{1.12} M_{Jup}$ and obtained discrepant day/night-side and equilibrium temperatures based on a claimed detection of the secondary eclipse. However, as we noted and  \cite{angerhausen14} concluded as well,  the detection of the secondary eclipse is neither clear nor conclusive. Moreover, given their derived mass for this companion with a Jupiter-like radius, it is difficult to explain their proposed stellar nature. These types of objects are only found in very young stellar associations (with ages younger than 10 Myr) such as Collinder 69 or $\sigma$-Orionis. Thus, even assuming their derived higher mass, this proposed false-positive configuration can be ruled out.

The alternative explanation provided by \cite{esteves13} suggests that the system is actually an eclipsing binary diluted by a contaminating third stellar source, a foreground or background star. However, \cite{lillo-box12,lillo-box14b} provided high-resolution images of this system (among other {\it Kepler} candidates) to rule out this and other configurations. 
According to Sect.~4.1.2 of the latter paper \citep[based on][]{law13}, the planetary transit of Kepler-91b cannot be mimicked by the presence of a diluted star \textit{fainter} than the transited star. Only a blended source with a very small radius that is \textit{brighter} than the transited star can dilute the binary eclipse and mimic a planetary transit. However, in the high-resolution image obtained by \cite{lillo-box14b} we did not find companions farther away than 0.1 arcsec. Thus, the probability for an undetected chance-aligned source brighter than 12.5 magnitudes and closer than 0.1 arcsec is lower than $10^{-6}$.

Sharing the opinion of \cite{esteves13}, \cite{sliski14} also classified Kepler-91b as a false positive. They compared the asteroseismic determination of the stellar density $\rho_{\star,astero}=6.81\pm0.32~kg/m^3$ \citep[derived by ][]{huber13}, with the observed value $\rho_{\star ,obs}=43.47^{+0.67}_{-3.35}~kg/m^3$ derived by them directly from transit fitting.  Based on the large discrepancy between the two values, they concluded that to explain this disagreement, the orbit of the planet would have to be highly eccentric such that the planet is essentially expected to be in-contact with the star. Basically, their derived observed stellar density corresponds to a semi-major axis of the companion of $ a/R_{\star}=4.476^{+0.023}_{-0.118}$, similar to that of \cite{esteves13}, $a/R_{\star}=4.5$. The authors claimed that the two determinations of the semi-major axis are independent, but they both used the same set of photometric data and the same observational effect (the transit signal).

Instead, we have determined the semi-major axis of the orbit by using three truly independent observational effects, namely the transit fitting $(a/R_{\star})_{transit} = 2.40\pm 0.12$ \citep[also supported by the previous analysis of][]{tenenbaum12}, light-curve modulations in the out-of-transit region  ($a/R_{\star})_{REB} = 2.45^{+0.15}_{-0.30}$, and the current radial velocity analysis ($a/R_{\star})_{RV} = 2.48\pm 0.12$. All three estimations provide autonomous and coincident measurements of the semi-major axis. As stated by \cite{sliski14} and already pointed out in \cite{lillo-box14}, this lower value implies an observed stellar density that agrees excellently with the asteroseismic analysis, thus confirming the planetary nature of the object orbiting Kepler-91 and rejecting the self-luminous scenario.  Possible explanations for the large discrepancy of the other determinations of $a/R_{\star}$ might involve some assumptions when fitting the transit. They both assumed i) a circular orbit, while we found that the shape of the ellipsoidal variations need a non-zero - although low- eccentricity, and ii) a spherical shape for the host star, which does not apply here since we clearly find light-curve modulations due to deformations of the stellar atmosphere.

The results presented here, together with the original confirmation paper \citep{lillo-box14}, provide strong evidence for the planetary-nature of Kepler-91b. Finally, this is the first planet confirmed from CAHA using CAFE, and we have proved the capability of this instrument for this type of research.

%-------------------------------------------------------------------

\begin{acknowledgements}
      This research has been funded by Spanish grant AYA2012-38897-C02-01. J. L-B thanks the CSIC JAE-predoc programme for the PhD fellowship. We also thank CAHA for allocating our observing runs. PF and NCS acknowledge support by  Funda\c{c}\~ao para a Ci\^encia e a Tecnologia (FCT) through Investigador FCT contracts of reference IF/01037/2013 and IF/00169/2012, respectively, and POPH/FSE (EC) by FEDER funding through the program ``Programa Operacional de Factores de Competitividade - COMPETE''. We also acknowledge the support from the European Research Council/European Community under the FP7 through Starting Grant agreement number 239953. 
\end{acknowledgements}

%______________________________________________________________

\bibliographystyle{aa} % style aa.bst
\bibliography{biblio2} % your references Yourfile.bib

\end{document}

%% file: Table_Observations.tex
\begin{table*}
\setlength{\extrarowheight}{2pt}
\scriptsize
\caption{Observational data and determined radial velocity. Julian date is calculated at mid-observation.}             
\label{tab:observations}      
\centering          
\begin{tabular}{c c c c | c c c c | c c c c }     % 7 columns 
\hline\hline       

 Julian Date       &    $\overline{S/N}$  &  Phase   &   RV              &  Julian Date       & $\overline{S/N}$  &  Phase   &   RV     &  Julian Date       & $\overline{S/N}$  &  Phase   &   RV       \\
 (days)-2456000                    &                      &          &   (km/s)          &  (days)-2456000                    &                   &          &   (km/s) &  (days)-2456000                    &                   &          &   (km/s)  \\ \hline

     079.3736990  &    10.2   &  0.6936  &   $-61.849^{+0.102}_{-0.086}$   &          095.6096755  &    10.7   &  0.2928  &   $-62.205^{+0.090}_{-0.075}$   &   116.4939625  &   11.3    &  0.6361  &   $-61.851^{+0.105}_{-0.086}$ \\
     079.4035263  &    10.9   &  0.6983  &   $-61.916^{+0.111}_{-0.096}$   &          095.6444425  &    10.5   &  0.2983  &   $-62.175^{+0.109}_{-0.094}$   &   121.3892272  &   10.8    &  0.4197  &   $-62.057^{+0.116}_{-0.096}$    \\     
     079.4383387  &    11.6   &  0.7039  &   $-61.905^{+0.099}_{-0.084}$   &          096.4018846  &    10.3   &  0.4196  &   $-62.051^{+0.157}_{-0.129}$   &   121.4241545  &   11.0    &  0.4253  &   $-62.129^{+0.115}_{-0.092}$    \\     
     080.6188492  &    10.0   &  0.8929  &   $-61.973^{+0.106}_{-0.090}$   &          096.4367742  &     9.6   &  0.4252  &   $-61.906^{+0.139}_{-0.115}$   &   123.5107198  &   10.8    &  0.7594  &   $-62.069^{+0.116}_{-0.097}$    \\     
     080.6408499  &     9.5   &  0.8964  &   $-62.006^{+0.095}_{-0.085}$   &          099.3819275  &    10.7   &  0.8966  &   $-62.062^{+0.115}_{-0.095}$   &   123.5616251  &   11.0    &  0.7675  &   $-61.968^{+0.100}_{-0.080}$    \\     
     089.6348116  &    11.3   &  0.3363  &   $-62.059^{+0.092}_{-0.078}$   &          099.4167349  &    10.5   &  0.9022  &   $-61.971^{+0.106}_{-0.091}$   &   128.3874072  &   10.5    &  0.5401  &   $-61.984^{+0.110}_{-0.090}$    \\  
     092.4534211  &    11.3   &  0.7875  &   $-61.933^{+0.091}_{-0.079}$   &          102.5440124  &   11.4    &  0.4029  &   $-61.994^{+0.082}_{-0.071}$   &   128.4226132  &   11.1    &  0.5457  &   $-62.014^{+0.105}_{-0.087}$  \\    
     092.4885156  &    11.5   &  0.7931  &   $-61.910^{+0.081}_{-0.073}$   &          102.5786588  &   11.6    &  0.4084  &   $-62.060^{+0.087}_{-0.073}$   &   131.3837431  &   11.4    &  0.0197  &   $-61.955^{+0.100}_{-0.086}$  \\    
     093.3820328  &    11.1   &  0.9361  &   $-61.984^{+0.092}_{-0.080}$   &          103.4294963  &   11.6    &  0.5446  &   $-61.942^{+0.085}_{-0.073}$   &   131.4186423  &   11.5    &  0.0253  &   $-61.993^{+0.099}_{-0.086}$  \\    
     093.4167639  &    11.4   &  0.9417  &   $-62.027^{+0.098}_{-0.080}$   &          103.4647180  &   10.6    &  0.5502  &   $-61.992^{+0.111}_{-0.089}$   &   140.3877688  &   11.0    &  0.4612  &   $-61.973^{+0.093}_{-0.077}$  \\    
     094.3825217  &    11.4   &  0.0963  &   $-62.023^{+0.092}_{-0.078}$   &          111.6216095  &   10.9    &  0.8561  &   $-62.030^{+0.100}_{-0.089}$   &   140.4227569  &   10.8    &  0.4668  &   $-61.895^{+0.121}_{-0.105}$  \\    
     094.4174484  &    11.6   &  0.1019  &   $-62.055^{+0.088}_{-0.077}$   &          112.3867977  &   11.4    &  0.9786  &   $-62.133^{+0.127}_{-0.105}$   &   141.5906808  &   10.6    &  0.6537  &   $-61.944^{+0.164}_{-0.134}$  \\          
     094.6184378  &    10.6   &  0.1341  &   $-62.018^{+0.081}_{-0.067}$   &          112.4215921  &   11.5    &  0.9841  &   $-62.023^{+0.124}_{-0.105}$   &    & & &  \\     
     094.6431432  &    10.5   &  0.1380  &   $-62.149^{+0.091}_{-0.080}$   &          116.4573650  &   11.1    &  0.6302  &   $-61.874^{+0.101}_{-0.081}$   &    & & &  \\     
  
\hline                  
\end{tabular}

\end{table*}

%% file: Table_RVresults.tex
\begin{table}[h]
\setlength{\extrarowheight}{3pt}
\scriptsize
\caption{ Best-fit and derived parameters from the RV analysis. \label{tab:rvresults} }
\centering
\begin{tabular}{rccc}
\hline\hline

Parameter    & Value                  & Units     & Assumptions\tablefootmark{a} \\  \hline
$K$          &   $93\pm17$            &  m/s     & e (REB), P (TR)     \\
$V_{sys}$    &    $-62.011\pm0.011$    &  km/s    & e (REB), P (TR)   \\
$\omega$     &   $250\pm100$          & deg.     & e (REB), P (TR)        \\
$M_p\sin{i}$ &   $1.01\pm0.18$        & $M_{\rm Jup}$ & e (REB), P (TR), $M_{\star}$ (AS)    \\ 
$a$          &   $0.0726\pm0.0019$    &  AU           & e (REB), P (TR), $M_{\star}$ (AS)    \\ \hline
$M_p$        &   $1.09\pm 0.20$        &  $M_{\rm Jup}$   & i (TR)   \\    
$\rho_p$     &   $0.41^{+0.13}_{-0.08}$       & $\rho_{\rm Jup}$ & $R_p$ (TR)  \\ 
$a/R_{\star}$ &   $2.48\pm 0.12$       &                &  $R_{\star}$ (AS)  \\
\hline
\hline

\end{tabular}
\tablefoot{Parameters below the horizontal line are derived from those above.
\tablefoottext{a}{Assumed parameters and the method used to determine them. Parameters: e (eccentricity), P (period), $M_{\star}$ (stellar mass), $i$ (orbital inclination), $R_p$ (planet radius), and $R_{\star}$ (stellar radius). Methods: AS = asteroseismology , TR = transit fitting , REB = light-curve modulations fitting. Assumed values are taken from \cite{lillo-box14}.}}
\end{table}

%% file: Kepler91b_RVconfirmation.bbl
\begin{thebibliography}{31}
\expandafter\ifx\csname natexlab\endcsname\relax\def\natexlab#1{#1}\fi

\bibitem[{{Aceituno} {et~al.}(2013){Aceituno}, {S{\'a}nchez}, {Grupp}, {Lillo},
  {Hern{\'a}n-Obispo}, {Benitez}, {Montoya}, {Thiele}, {Pedraz}, {Barrado},
  {Dreizler}, \& {Bean}}]{aceituno13}
{Aceituno}, J., {S{\'a}nchez}, S.~F., {Grupp}, F., {et~al.} 2013, \aap, 552,
  A31

\bibitem[{{Angerhausen} {et~al.}(2014){Angerhausen}, {DeLarme}, \&
  {Morse}}]{angerhausen14}
{Angerhausen}, D., {DeLarme}, E., \& {Morse}, J.~A. 2014, ArXiv e-prints,
  1404.4348

\bibitem[{{Batalha} {et~al.}(2013){Batalha}, {Rowe}, {Bryson}, {Barclay},
  {Burke}, {Caldwell}, {Christiansen}, {Mullally}, {Thompson}, {Brown},
  {Dupree}, {Fabrycky}, {Ford}, {Fortney}, {Gilliland}, {Isaacson}, {Latham},
  {Marcy}, {Quinn}, {Ragozzine}, {Shporer}, {Borucki}, {Ciardi}, {Gautier},
  {Haas}, {Jenkins}, {Koch}, {Lissauer}, {Rapin}, {Basri}, {Boss}, {Buchhave},
  {Carter}, {Charbonneau}, {Christensen-Dalsgaard}, {Clarke}, {Cochran},
  {Demory}, {Desert}, {Devore}, {Doyle}, {Esquerdo}, {Everett}, {Fressin},
  {Geary}, {Girouard}, {Gould}, {Hall}, {Holman}, {Howard}, {Howell},
  {Ibrahim}, {Kinemuchi}, {Kjeldsen}, {Klaus}, {Li}, {Lucas}, {Meibom},
  {Morris}, {Pr{\v s}a}, {Quintana}, {Sanderfer}, {Sasselov}, {Seader},
  {Smith}, {Steffen}, {Still}, {Stumpe}, {Tarter}, {Tenenbaum}, {Torres},
  {Twicken}, {Uddin}, {Van Cleve}, {Walkowicz}, \& {Welsh}}]{batalha13}
{Batalha}, N.~M., {Rowe}, J.~F., {Bryson}, S.~T., {et~al.} 2013, \apjs, 204, 24

\bibitem[{{Borucki} {et~al.}(2009){Borucki}, {Koch}, {Jenkins}, {Sasselov},
  {Gilliland}, {Batalha}, {Latham}, {Caldwell}, {Basri}, {Brown},
  {Christensen-Dalsgaard}, {Cochran}, {DeVore}, {Dunham}, {Dupree}, {Gautier},
  {Geary}, {Gould}, {Howell}, {Kjeldsen}, {Lissauer}, {Marcy}, {Meibom},
  {Morrison}, \& {Tarter}}]{borucki09}
{Borucki}, W.~J., {Koch}, D., {Jenkins}, J., {et~al.} 2009, Science, 325, 709

\bibitem[{{Borucki} {et~al.}(2011){Borucki}, {Koch}, {Basri}, {Batalha},
  {Brown}, {Bryson}, {Caldwell}, {Christensen-Dalsgaard}, {Cochran}, {DeVore},
  {Dunham}, {Gautier}, {Geary}, {Gilliland}, {Gould}, {Howell}, {Jenkins},
  {Latham}, {Lissauer}, {Marcy}, {Rowe}, {Sasselov}, {Boss}, {Charbonneau},
  {Ciardi}, {Doyle}, {Dupree}, {Ford}, {Fortney}, {Holman}, {Seager},
  {Steffen}, {Tarter}, {Welsh}, {Allen}, {Buchhave}, {Christiansen}, {Clarke},
  {Das}, {D{\'e}sert}, {Endl}, {Fabrycky}, {Fressin}, {Haas}, {Horch},
  {Howard}, {Isaacson}, {Kjeldsen}, {Kolodziejczak}, {Kulesa}, {Li}, {Lucas},
  {Machalek}, {McCarthy}, {MacQueen}, {Meibom}, {Miquel}, {Prsa}, {Quinn},
  {Quintana}, {Ragozzine}, {Sherry}, {Shporer}, {Tenenbaum}, {Torres},
  {Twicken}, {Van Cleve}, {Walkowicz}, {Witteborn}, \& {Still}}]{borucki11}
{Borucki}, W.~J., {Koch}, D.~G., {Basri}, G., {et~al.} 2011, \apj, 736, 19

\bibitem[{{Coelho} {et~al.}(2005){Coelho}, {Barbuy}, {Mel{\'e}ndez},
  {Schiavon}, \& {Castilho}}]{coelho05}
{Coelho}, P., {Barbuy}, B., {Mel{\'e}ndez}, J., {Schiavon}, R.~P., \&
  {Castilho}, B.~V. 2005, \aap, 443, 735

\bibitem[{{Debosscher} {et~al.}(2007){Debosscher}, {Sarro}, {Aerts}, {Cuypers},
  {Vandenbussche}, {Garrido}, \& {Solano}}]{debosscher07}
{Debosscher}, J., {Sarro}, L.~M., {Aerts}, C., {et~al.} 2007, \aap, 475, 1159

\bibitem[{{Esteves} {et~al.}(2013){Esteves}, {De Mooij}, \&
  {Jayawardhana}}]{esteves13}
{Esteves}, L.~J., {De Mooij}, E.~J.~W., \& {Jayawardhana}, R. 2013, \apj, 772,
  51

\bibitem[{{Eyer} \& {Bartholdi}(1999)}]{eyer99}
{Eyer}, L. \& {Bartholdi}, P. 1999, \aaps, 135, 1

\bibitem[{{Faigler} \& {Mazeh}(2014)}]{faigler14}
{Faigler}, S. \& {Mazeh}, T. 2014, ArXiv e-prints, 1407.2361

\bibitem[{{Gould} \& {Loeb}(1992)}]{gould92}
{Gould}, A. \& {Loeb}, A. 1992, \apj, 396, 104

\bibitem[{{Holman} {et~al.}(2010){Holman}, {Fabrycky}, {Ragozzine}, {Ford},
  {Steffen}, {Welsh}, {Lissauer}, {Latham}, {Marcy}, {Walkowicz}, {Batalha},
  {Jenkins}, {Rowe}, {Cochran}, {Fressin}, {Torres}, {Buchhave}, {Sasselov},
  {Borucki}, {Koch}, {Basri}, {Brown}, {Caldwell}, {Charbonneau}, {Dunham},
  {Gautier}, {Geary}, {Gilliland}, {Haas}, {Howell}, {Ciardi}, {Endl},
  {Fischer}, {F{\"u}r{\'e}sz}, {Hartman}, {Isaacson}, {Johnson}, {MacQueen},
  {Moorhead}, {Morehead}, \& {Orosz}}]{holman10}
{Holman}, M.~J., {Fabrycky}, D.~C., {Ragozzine}, D., {et~al.} 2010, Science,
  330, 51

\bibitem[{{Huber} {et~al.}(2013){Huber}, {Chaplin}, {Christensen-Dalsgaard},
  {Gilliland}, {Kjeldsen}, {Buchhave}, {Fischer}, {Lissauer}, {Rowe},
  {Sanchis-Ojeda}, {Basu}, {Handberg}, {Hekker}, {Howard}, {Isaacson},
  {Karoff}, {Latham}, {Lund}, {Lundkvist}, {Marcy}, {Miglio}, {Silva Aguirre},
  {Stello}, {Arentoft}, {Barclay}, {Bedding}, {Burke}, {Christiansen},
  {Elsworth}, {Haas}, {Kawaler}, {Metcalfe}, {Mullally}, \&
  {Thompson}}]{huber13}
{Huber}, D., {Chaplin}, W.~J., {Christensen-Dalsgaard}, J., {et~al.} 2013,
  \apj, 767, 127

\bibitem[{{Ivezi{\'c}} {et~al.}(2013){Ivezi{\'c}}, {Connolly}, {VanderPlas}, \&
  {Gray}}]{ivezic13}
{Ivezi{\'c}}, {\.Z}., {Connolly}, A., {VanderPlas}, J., \& {Gray}, A. 2013,
  {Statistics, Data Mining, and Machine Learning in Astronomy} (Princeton
  University Press)

\bibitem[{{Law} {et~al.}(2013){Law}, {Morton}, {Baranec}, {Riddle},
  {Ravichandran}, {Ziegler}, {Johnson}, {Tendulkar}, {Bui}, {Burse}, {Das},
  {Dekany}, {Kulkarni}, {Punnadi}, \& {Ramaprakash}}]{law13}
{Law}, N.~M., {Morton}, T., {Baranec}, C., {et~al.} 2013, ArXiv e-prints,
  1312.4958

\bibitem[{{Lazorenko} {et~al.}(2011){Lazorenko}, {Sahlmann}, {S{\'e}gransan},
  {Figueira}, {Lovis}, {Martin}, {Mayor}, {Pepe}, {Queloz}, {Rodler}, {Santos},
  \& {Udry}}]{lazorenko11}
{Lazorenko}, P.~F., {Sahlmann}, J., {S{\'e}gransan}, D., {et~al.} 2011, \aap,
  527, A25

\bibitem[{{Lillo-Box} {et~al.}(2012){Lillo-Box}, {Barrado}, \&
  {Bouy}}]{lillo-box12}
{Lillo-Box}, J., {Barrado}, D., \& {Bouy}, H. 2012, \aap, 546, A10

\bibitem[{{Lillo-Box} {et~al.}(2014{\natexlab{a}}){Lillo-Box}, {Barrado}, \&
  {Bouy}}]{lillo-box14b}
{Lillo-Box}, J., {Barrado}, D., \& {Bouy}, H. 2014{\natexlab{a}}, \aap, 566,
  A103

\bibitem[{{Lillo-Box} {et~al.}(2014{\natexlab{b}}){Lillo-Box}, {Barrado},
  {Moya}, {Montesinos}, {Montalb{\'a}n}, {Bayo}, {Barbieri}, {R{\'e}gulo},
  {Mancini}, {Bouy}, \& {Henning}}]{lillo-box14}
{Lillo-Box}, J., {Barrado}, D., {Moya}, A., {et~al.} 2014{\natexlab{b}}, \aap,
  562, A109

\bibitem[{{Mao} \& {Paczynski}(1991)}]{mao91}
{Mao}, S. \& {Paczynski}, B. 1991, \apjl, 374, L37

\bibitem[{{Mayor} \& {Queloz}(1995)}]{mayor95}
{Mayor}, M. \& {Queloz}, D. 1995, \nat, 378, 355

\bibitem[{{Muterspaugh} {et~al.}(2010){Muterspaugh}, {Lane}, {Kulkarni},
  {Konacki}, {Burke}, {Colavita}, {Shao}, {Hartkopf}, {Boss}, \&
  {Williamson}}]{muterspaugh10}
{Muterspaugh}, M.~W., {Lane}, B.~F., {Kulkarni}, S.~R., {et~al.} 2010, \aj,
  140, 1657

\bibitem[{{Quintana} {et~al.}(2013){Quintana}, {Rowe}, {Barclay}, {Howell},
  {Ciardi}, {Demory}, {Caldwell}, {Borucki}, {Christiansen}, {Jenkins},
  {Klaus}, {Fulton}, {Morris}, {Sanderfer}, {Shporer}, {Smith}, {Still}, \&
  {Thompson}}]{quintana13}
{Quintana}, E.~V., {Rowe}, J.~F., {Barclay}, T., {et~al.} 2013, \apj, 767, 137

\bibitem[{{Santerne} {et~al.}(2012){Santerne}, {D{\'{\i}}az}, {Moutou},
  {Bouchy}, {H{\'e}brard}, {Almenara}, {Bonomo}, {Deleuil}, \&
  {Santos}}]{santerne12}
{Santerne}, A., {D{\'{\i}}az}, R.~F., {Moutou}, C., {et~al.} 2012, \aap, 545,
  A76

\bibitem[{{Shporer} {et~al.}(2011){Shporer}, {Jenkins}, {Rowe}, {Sanderfer},
  {Seader}, {Smith}, {Still}, {Thompson}, {Twicken}, \& {Welsh}}]{shporer11}
{Shporer}, A., {Jenkins}, J.~M., {Rowe}, J.~F., {et~al.} 2011, \aj, 142, 195

\bibitem[{{Sliski} \& {Kipping}(2014)}]{sliski14}
{Sliski}, D.~H. \& {Kipping}, D.~M. 2014, ArXiv e-prints, 1401.1207

\bibitem[{{Tenenbaum} {et~al.}(2012){Tenenbaum}, {Jenkins}, {Seader}, {Burke},
  {Christiansen}, {Rowe}, {Caldwell}, {Clarke}, {Li}, {Quintana}, {Smith},
  {Thompson}, {Twicken}, {Borucki}, {Batalha}, {Cote}, {Haas}, {Sanderfer},
  {Girouard}, {Hall}, {Ibrahim}, {Klaus}, {McCauliff}, {Middour}, {Sabale},
  {Kamal Uddin}, {Wohler}, {Barclay}, \& {Still}}]{tenenbaum12}
{Tenenbaum}, P., {Jenkins}, J.~M., {Seader}, S., {et~al.} 2012, ArXiv e-prints,
  1212.2915

\bibitem[{{Vanderplas} {et~al.}(2012){Vanderplas}, {Connolly}, {Ivezi{\'c}}, \&
  {Gray}}]{vanderplas12}
{Vanderplas}, J., {Connolly}, A., {Ivezi{\'c}}, {\v Z}., \& {Gray}, A. 2012, in
  Conference on Intelligent Data Understanding (CIDU), 47 --54

\bibitem[{{Wang} {et~al.}(2012){Wang}, {Wright}, {Cochran}, {Kane}, {Henry},
  {Payne}, {Endl}, {MacQueen}, {Valenti}, {Antoci}, {Dragomir}, {Matthews},
  {Howard}, {Marcy}, {Isaacson}, {Ford}, {Mahadevan}, \& {von Braun}}]{wang12}
{Wang}, Sharon, X., {Wright}, J.~T., {Cochran}, W., {et~al.} 2012, \apj, 761,
  46

\bibitem[{{Wright} \& {Howard}(2009)}]{wright09}
{Wright}, J.~T. \& {Howard}, A.~W. 2009, \apjs, 182, 205

\bibitem[{{Zucker}(2003)}]{zucker03}
{Zucker}, S. 2003, \mnras, 342, 1291

\end{thebibliography}
